\documentstyle[epsfig]{elsart}
\begin{document}
\begin{frontmatter}
\title{Single Event Effects in the Pixel readout chip for BTeV}
\author{G. Chiodini},
\author{J.A. Appel},
\author{G. Cardoso}, 
\author{D.C. Christian},
\author{M.R. Coluccia}, 
\author{J. Hoff}, 
\author{S.W. Kwan}, 
\author{A. Mekkaoui},
\author{R. Yarema}, and
\author{S. Zimmermann}
\address{Fermi National Accelerator Laboratory, Batavia, IL 60510, USA}

\begin{abstract}
In future experiments the readout electronics 
for pixel detectors is required to be resistant to 
a very high radiation level. 
In this paper we report on irradiation tests
performed on several preFPIX2 prototype pixel 
readout chips for the BTeV experiment
exposed to a 200 MeV proton beam.
The prototype chips have been implemented in 
commercial 0.25 $\mu$m CMOS processes following 
radiation tolerant design rules.
The results show that this ASIC design tolerates
a large total radiation dose, and that radiation
induced Single Event Effects occur at a manageable level.
\end{abstract}
\end{frontmatter}

\section{Introduction}
Several future high energy physics experiments 
(ALICE, ATLAS, BTeV, and CMS) are planning to use 
vertex detectors based on hybrid silicon pixel detectors. 
These detectors will experience a large total ionizing dose 
and neutron flux. Consequently, the front end electronics,
covering entirely the active area of the sensors,
must be very radiation tolerant.

The front end electronics for hybrid pixel detetectors
is realized in CMOS technology in order to keep
the amount of power dissipation to a manageable
level (about 0.5 W$\cdot$cm$^{-2}$). 
CMOS devices are sensitive to ionizing radiation due
to the positive charge-up of the silicon oxide. 
The trapped charge causes a voltage threshold shift 
(charge-up of gate oxide), 
leakage current increases within NMOS devices
(charge-up of the oxide surrounding the device)
and between NMOS devices (charge-up of the oxide separating devices). 
The radiation damage can be greatly reduced by
using commercial deep-submicron technology for ASIC design.
In these technologies the 
gate oxide thickness is so small (t$_{ox} < $ 6 nm)
that electron tunneling is extremely effective
in removing trapped holes and stopping the formation
of interface states. 
Moreover, even in deep-submicron technology,
the field oxide is relatively thick. This means that 
the radiation-induced leakage currents must be prevented 
by enclosed-gate NMOS devices (no leakage currents between drain and
source) and NMOS devices with guard rings 
(no leakage currents between NMOS devices).
Following these layout rules \cite{Adams} the radiation hardness
to total dose is found to increase enormously.
However, Total Ionizing Dose  (TID) effects
are not the only concern in
a radiation enviroment. Ionization radiation can  
deposit enough energy density by recoils from nuclear 
interaction to cause Single Event Effects (SEE).      
There are three important SEE in CMOS technology
to be considered: Single Event Latch-up (SEL), 
Single Event Gate Rupture (SEGR), and Single Event Upset (SEU).  

In SEL, a device which is supposed to be off 
(or a parasitic device
in parallel to the real one) 
is turned on, providing a path from
the supply to ground. A single latch-up can result in a
large current and destruction of the IC or melting of
wire bonds providing power if fast current-limiting 
protection is not in place.
SEL sensitivity has been observed to decrease in deep 
submicron processes for a variety of reasons \cite{Johnston} 
(reduced thickness of the epitaxial layer, retrograde wells,
and Shallow Trench Isolation). In addition, the use of
guard rings around NMOS devices helps to prevent latch-up.

In SEGR total or partial damage of
the dielectric gate material occurs
due to an  avalanche discharge.
In reference \cite{Sexton} studies of the dependence 
of SEGR on gate oxide thickness show that the phenomenon
is likely not to be a concern in deep-submicron technology.
The Critical Field (E$_{c}$) to rupture, for a given Linear Energy
Transfer (LET), increases for decreasing gate oxide thickness.
In particularly, for t$_{ox} < $ 6 nm, the critical field is higher
than 7 MeV$\cdot$cm$^{-1}$ for LET $<$ 80 MeV$\cdot$cm$^{2}$mg$^{-1}$,
this is significantly larger than the electric field present
in dielectric gate oxide in 0.25 $\mu$m CMOS devices.   

In SEU a soft error is introduced in logic circuits due
to an electric glitch positively 
amplified by the circuit itself. 
The phenomenon is not a distructive one, but it can 
change the state of a flip-flop or induce other
unwanted logic state transitions.
SEU effects are possible if the local energy deposited, 
or equivalently the local charge deposited, is large enough. 
The critical energy depends strongly on the technology used. 
In particular, a technology with smaller feature size generally
has smaller critical energy. 
This doesn't necessary mean that deep-submicron technology
is more prone to soft errors. In fact, to cause upset,
the charge must be released near a sensitive node, 
(often the drain of a CMOS device). 
The sensitive volumes are smaller for small feature size
technology than for larger feature size
technology. 
There are circuit hardening techniques to mitigate
single event upset \cite{Faccio}, 
but usually these require
an increase of the circuit area, complexity, and power    
consumptions - all factors already constrained in 
a readout chip for pixel sensors.  

In conclusion, modern deep-submicron CMOS processes are not expected
to suffer from SEL and SEGR effects, but can be quite sensitive
to SEU effects. For this reason, 
we did proton irradiation tests with BTeV readout chip 
prototypes in order to verify the robustness 
to total dose and Single Event Effects.  
 
\section{The FPIX pixel readout chip for BTeV}
\subsection{Introduction}
The BTeV experiment plans to run at the Tevatron Collider in 2006 [1].
It is designed to cover the ``forward'' regions of the proton-antiproton 
interaction point running at a luminosity of 2$\cdot$10$^{32}$ cm$^{-2}$s$^{-1}$. 
The experiment will employ a silicon pixel 
vertex detector \cite{Marina} to provide high precision space 
points for an on-line lowest-level trigger 
based on track impact parameters \cite{Wang}. 
The ``hottest'' chips, located at 6 mm from the beam, 
will experience a fluence of about 10$^{14}$ cm$^{-2}$y$^{-1}$. 
This is similar to the high radiation environments at ATLAS and CMS at LHC.

\begin{figure}
\begin{center}
\vspace{30mm}
\epsfig{figure=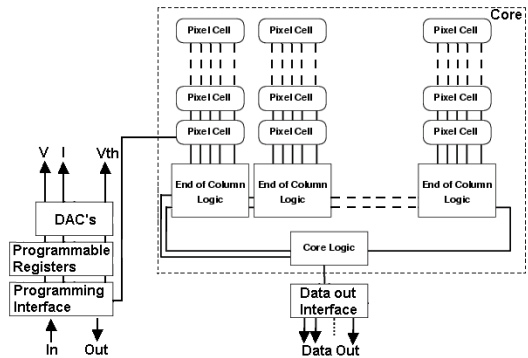,width=70mm,angle=0}
\end{center}
\caption{FPIX pixel readout chip block diagram.
\label{coreper} }
\end{figure}

In order to satisfy the needs of BTeV, the FPIX pixel readout chip must 
provide ``very clean'' track crossing information near the interaction region 
for every 132 ns beam crossing. This requires a low noise front-end, 
an unusually high output bandwidth, and radiation-hard technology.
A pixel detector readout chip (FPIX) has been developed at 
Fermilab to meet the requirements of future Tevatron Collider experiments.
In Figure~\ref{coreper} the FPIX chip block diagram is shown. 
It consists of an array of 22 columns of 128 cells of 
50 $\mu$m by 400 $\mu$m pixel electronic cells. 
Each front end cell implemements a DC leakage current 
compensation circuit and a 3 bit flash ADC. 

The readout of a column is by end of column logic
through a cell token-pass signal. 
In the end of column, up to four timestamps can be stored 
and associated with the corresponding cells over threshold.   
The readout of the columns is arbitrated by the core logic
through a column token-pass signal and the data are 
sent off chip by a fast data output serializer. 
The pixel mask and charge-injection registers, and 
digital-to-analog (DAC) registers are downloaded to the FPIX
through a serial programming interface.
The mask and charge-injection registers consist of small 
size daisy chained flip-flop's (FF's) and are implemented in 
each pixel cell. 
A high logic level stored in one of the mask FF's disables 
the corresponding cell. 
This is meant to turn off noisy cells. 
Analogously, a high logic level stored in one of the 
charge-injection FF's enables the cell to receive at the 
input an analogue pulse for calibration purposes. 
Thus, there are two independent long registers, 
which are serpentine through the chip. 
The DAC registers control features of the chip and minimize 
the number of connections between the chip and the outside world.
In the DAC's the stored digital value is translated into an 
analogue voltage or analogue current to set bias voltages, 
bias currents and  discriminator thresholds. 
\subsection{The preFPIX2 chip prototypes}
The preFPIX2 represents the most advanced member of very 
successful succession of pixel readout chip prototypes \cite{Christian}.
It has been realized in standard 0.25$\mu$m CMOS technology, 
following radiation tolerant design rules, from
two vendors \cite{Mekkaoui}.
Last year exposures of preFPIX2T chips to radiation from a Colbalt-60 
source at Argonne National Laboratory verified the high tolerance 
to gamma radiation up to about 33 MRad total dose \cite{Mekkaoui}. 
In this paper, we present results of proton radiation tests performed with 
preFPIX2 chip prototypes including both total dose and single event effects.

The proton irradiation tests were performed exposing two
chip prototypes, preFPIX2I and preFPIX2Tb chip, to 200 MeV proton beam 
at the Indiana University Cyclotron Facility (IUCF). 
The preFPIX2I chip, containing 16 columns with 32 rows of pixel cells 
and complete core readout architecture, was manufactured by a
vendor through CERN. 
The preFPIX2Tb chip contains, in addition to the preFPIX2I chip features, 
a new programming interface and 14 8-bit DAC's. 
It was manufactured by Taiwan Semiconductor Manufacturing Company. 
The comparison of the chip performance before and after exposure shows the high 
radiation tolerance of the design to protons up to about 43 MRad total dose. 
In addition, we measured the SEU cross section of static registers
implemented on the preFPIX2Tb chip, in order to establish the sensitivity
of our design to radiation induced digital soft errors during real operation.
\section{Experimental setup}
\subsection{Proton irradiation facility at IUCF}
The proton irradiation tests took place at the 
Indiana University Cyclotron Facility where a proton 
beam line of 200 MeV kinetic energy is delivered to users. 
The beam profile has been measured by exposing a photographic film. 
The beam spot, defined by the circular area where 
the flux is not less than 90\% of the central value, 
had a diameter of about 15 mm, comfortably larger than the chip size 
(the larger chip is preFPIX2Tb which is 4.3 mm wide and 7.2 mm long). 
Before the exposure the absolute fluence was measured by a Faraday cup; 
during the exposure by a secondary electron emission monitor. 
The cyclotron has a duty cycle factor of 0.7\% with a repetition 
rate of about 17MHz and most of the tests were done with a flux 
of about 2$\cdot$10$^{10}$ protons cm$^{-2}$s$^{-1}$.   
The irradiation was done with the chips powered on,
in air at room temperature, and no low energy particle 
or neutron filters were used. 
The exposures with multiple boards were done placing the 
boards about 2 cm behind each other and with the chips facing the beam. 
Mechanically, the boards were kept in position by an open aluminium frame. 
The beam was centred on the chips. The physical position of the 
frame was monitored constantly by a video camera to ensure that no 
movements occurred during exposure.  
We irradiated 4 boards with preFPIXI chips to 26 MRad (December 2000), 
one board with preFPIX2Tb to 14 MRad (April 2001), and 
4 boards with preFPIX2Tb to 29 MRad (August 2001). One of the boards
with preFPIX2Tb chips on it was irradiated twice, collecting 43 MRad total dose. 
Due to the alignment precision and measurement technique employed, 
the systematic error on the integrated fluence is believed to be less than 10\%.

\subsubsection{LET spectra of 200 MeV protons in silicon}
In reference \cite{O'Neil} the production of ions in bulk silicon by a 
200 MeV proton beam is described by a Monte Carlo simulation as a two
stage process. In the first stage (internuclear cascade) the proton hits 
the silicon nucleus and produces light fragments, mainly forward. 
In the second stage the struck heavy ion (from nitrogen to silicon) 
recoils and evaporates isotropically producing further light fragments.
In these inelastic collisions, the light fragments produced
in the internuclear cascade, have a long range 
(up to more than 100 $\mu$m) and low lineary energy transfer
(less than 1.5 MeV$\cdot$cm$^{2}$mg$^{-1}$). 
The recoiling heavy ions 
have a shorter range (about 10 $\mu$m) and relatively high 
lineary energy transfer (up to
14 MeV$\cdot$cm$^{2}$mg$^{-1}$) and can be very effective in
causing SEU in electronics. 
In reference \cite{Jarron} the lineary energy transfer 
threshold for SEU in a static 
register realized in deep-submicron 
technology has been measured using 
ion beams of various species (i.e., various lineary energy transfer
values)
to be 14.7 MeV$\cdot$cm$^{2}$mg$^{-1}$.
This value is quite high for a 0.25 $\mu$m technology, but
small enough to expect an SEU sensitivity to 200 MeV protons.

\subsection{Hardware and software DAQ}
Each chip under test was wire-bonded to a printed circuit board 
in such a way that it could be properly biased,
controlled and read out by a DAQ system. 
The DAQ system was based on a PCI card designed at 
Fermilab (PCI Test Adapter Card) plugged in a PCIbus 
extender and controlled by a laptop PC. 
The PTA card generated digital signals to 
control and read back the readout chips.  
The software to control the PCI card IO busses 
was custom written in C. 
The PCI card busses were buffered by LVDS 
differential driver-receiver cards near the 
PCIbus extender located in the counting room. 
The differential card drove a 100 foot twisted pair 
cable followed by another LVDS differential driver-receiver 
card which finally was connected with a 10 foot flat 
cable to the devices under test.  
All the DAQ electronics were well behind thick concrete walls, 
protecting the apparatus from being influenced by the radiation 
background from the cyclotron and from activated material.        
\section{Experimental results}
\subsection{Single Event Latch-up}
During the irradiation tests
the circuits were powered with an applied bias voltage
of 2.5V. The analogue and digital 
currents were continuously monitored via GPIB. 
The analogue current decreased slightly and the digital currents 
increased slightly during the proton exposure.
No power supply trip-offs or large increases in the bias currents 
were observed during the irradiation. 
There is no evidence of single event latch-up or of significant 
radiation induced leakage currents.
This result is expected from reference \cite{Jarron} where
no evidence of latch-up up to linear energy transfer
of 89 MeV$\cdot$cm$^{2}$mg$^{-1}$
was reported.
\subsection{Single Event Gate Rupture}
The damage of a CMOS device due to a  
Single Event Gate Rupture doesn't necessary mean 
a hard failure of the chip but is likely 
to show up as performance degradation.
In order to detect the occurance of SEGR due
to the proton exposure, we measured the noise and 
the discriminator threshold of each individual cell 
before and after irradiation. 
The presence after exposure of a pixel cell 
with a noise or threshold significantly different 
from the other ones is an indication
of radiation damage localized only
in those particular cells, possibly the occurance of 
SEGR of individual transistors.
So far we have completed the mentioned test
for all four preFPIX2I boards, irradiated
up to 26 MRad, and one of the preFPIX2Tb board, 
irradiated up to 43 MRad.
We screened a total of about 2300 pixel front-end
cells, each one containing about 550 transistors.
All the cells were working before and after 
the irradiation, and none of them showed
peculiar behavior.
What has been observed after the irradiation
has been a slight increase of the average threshold
that can be adjusted by changing an internal bias-voltage 
reference (details about the FPIX2 front-end 
characterization after gamma and proton irradiation
can be found in reference \cite{Hoff}). 
 
\begin{figure}
\begin{center}
\vspace{30mm}
\epsfig{figure=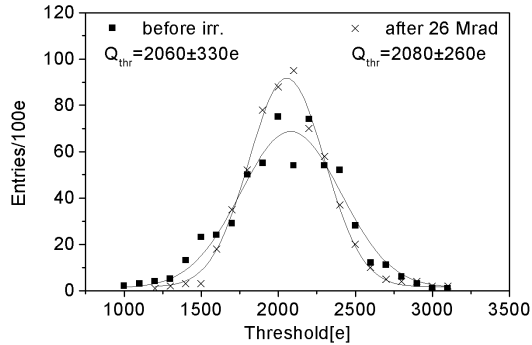,width=70mm,angle=0}
\end{center}
\caption{Measured discriminator threshold in the 576 cells of preFPIX2I 
before and after 26 MRad of 200 MeV proton irradiation 
(Note: the $\pm$ values are widths of the fitted Gaussian curves).
\label{threshold26mrad} }
\end{figure}

\begin{figure}
\begin{center}
\vspace{30mm}
\epsfig{figure=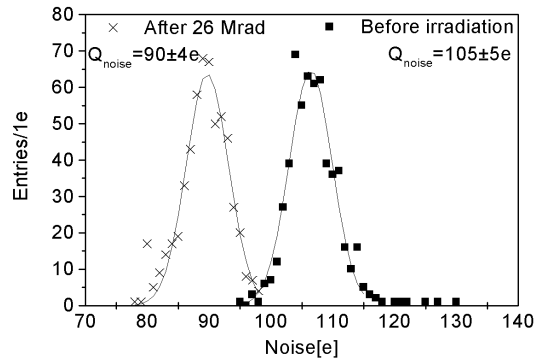,width=70mm,angle=0}
\end{center}
\caption{Measured amplifier noise in the 576 cells of preFPIX2I 
before and after 26 MRad of 200 MeV proton irradiation
(Note: the $\pm$ values are widths of the fitted Gaussian curves).
\label{noise26mrad} }
\end{figure}

\begin{figure}
\begin{center}
\vspace{30mm}
\epsfig{figure=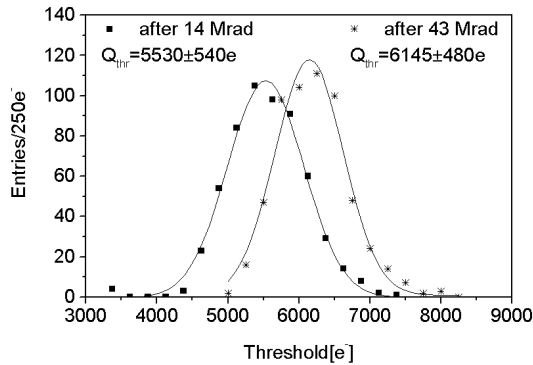,width=70mm,angle=0}
\end{center}
\caption{Measured discriminator threshold in the 576 cells of preFPIX2Tb 
after 14 and 43 MRad of 200 MeV proton irradiation
(Note: the $\pm$ values are widths of the fitted Gaussian curves).
\label{threshold43mrad} }
\end{figure}

\begin{figure}
\begin{center}
\vspace{30mm}
\epsfig{figure=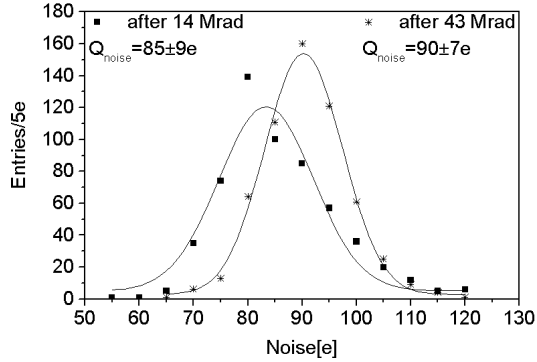,width=70mm,angle=0}
\end{center}
\caption{Measured amplifier noise in the 576 cells of preFPIX2Tb 
after 14 and 43 MRad of 200 MeV proton irradiation
(Note: the $\pm$ values are widths of the fitted Gaussian curves).
\label{noise43mrad} }
\end{figure}

Figures ~\ref{threshold26mrad} and ~\ref{noise26mrad}
show the threshold and noise distributions 
of a preFPIXI chip irradiated with a proton dose of 26 MRad. 
For these measurements the chip was biased exactly 
the same before and after irradiation except for the reference
voltage of the second stage amplifier which
was adjustted in order to have the
same average threshold.
Figures ~\ref{threshold43mrad} and ~\ref{noise43mrad}
show the threshold and noise distributions 
of preFPIXTb chips irradiated with a proton dose of 14 and
43 MRad. For these measurements, the chips were downloaded 
with the same DAC values and a shift in the average threshold 
is clearly visible between 14 and 43 MRad total proton dose.
The absence of noisy cells, and of large differences 
in individual thresholds due to irradiation, 
strongly suggest that single event gate rupture is not a concern
in this design.
\subsection{DAC analog response}
\begin{figure}
\begin{center}
\vspace{30mm}
\epsfig{figure=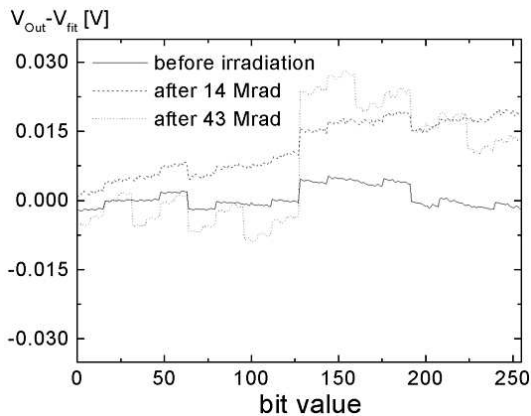,width=70mm,angle=0}
\end{center}
\caption{DAC analog responce before and after 14 and 43 MRad total dose exposure to
 200 MeV protons. The full scale (255 counts) corresponds to about 1.7 V.
\label{Vth0diff} }
\end{figure}

The analog output of the DAC's implemented in preFPIX2Tb 
are available external to the chip.
This means that change of the DAC behavior 
due to the proton irradiation can be measured.
Figure ~\ref{Vth0diff} shows
the response of the DAC used to adjust 
the discriminator thresholds before and 
after proton irradiation.
The three curves shown
correspond to the deviation from the linear fit
for total doses of 0, 14, and 43 MRad.
It is possible to see that the linearity and
accuracy of the DAC output remain acceptable
after 43 MRad total dose. 
The highest deviation from the linear fit 
is seen when the most significant bit
is set true. In fact, the most significant
bit has an associated current source drawing 
the same amount of current as all the other bits
together, and a mismatch induced by the
radiation shows up more in percentage.
The spread could be
quantified in terms of number of counts, and turn out
that in the extreme situation of a total dose
of 43 MRad, the digit inaccuracy is not more than 4 counts,
where 255 counts is full scale.
The behavior of the other DAC's for high 
total dose is very similar to the one shown in Figure ~\ref{Vth0diff}. 
\subsection{Single Event Upsets}
In our tests, a great deal of attention was focused on measuring 
radiation-induced digital soft errors.  
We concentrated our effort on the preFPIX2Tb registers 
storing the initialization parameters, 
because they have a large number of bits and the testing 
procedure is easy to perform. 
The results obtained allow prediction of the
performance of other parts of the chip potentially 
affected by the same phenomena. 
\subsubsection{Testing procedure}
The single event upset tests performed are very similar to 
the ones reported in reference \cite{Jarron}. 
The SEU measurements consisted of detecting single bit 
errors in the values stored in the registers. 
The testing procedure consisted of repeatedly downloading 
all the registers and reading back the stored values after one minute. 
The download and read-back phases took about 3 seconds. 
The download of the parameters was done with a pattern 
with half of the stored bits having a logical value 0 and 
the other half having a logical value 1 
(except in one case, see the caption in Table ~\ref{seuerror}).
For the shift-registers, the patterns were 
randomly generated at every iteration loop. 
For the DAC registers, the patterns were kept constant. 
A mismatch between the read-back value and the download value 
is interpreted as a single event upset due to the proton irradiation. 
No errors were observed in the system with the beam off 
and running for 10 hours.

In a separate test, the mask register of one board was 
operated in clocked mode with a clock frequency of 380 kHz. 
The low clock frequency value was due to our DAQ limitation. 
In this test, the mask register was downloaded with a 
logical level 1 in each flip-flop, in order to increase the 
statistics in view of the fact that a stored logical 
level 1 is easier to upset than a logical level 0 (see results). 
After the initialization, a continuous read cycle was 
performed and stopped every time a logical level 0 was detected. 
We collected 14 errors during an integrated fluence 
of 5.8$\cdot$10$^{13}$ protons cm$^{-2}$.

A summary of the total single bit errors detected in the
preFPIX2Tb readout chips, together with other relevant quantities, 
is shown in Table ~\ref{seuerror}.
\begin{table}
\begin{center}
\begin{tabular}{|l|l|l|l|} \hline
Chip           &  Int. Fluence [$cm^{-2}]$ & Shift-Reg errors		      & DAC-Reg errors\\ \hline
1              &  2.33$\cdot10^{14}      $ &53=18($\uparrow$)+35($\downarrow$) & 10=8 ($\uparrow$)+2($\downarrow$) \\ \hline
2              &  3.65$\cdot10^{14}      $ &74=22($\uparrow$)+52($\downarrow$) & 19=9 ($\uparrow$)+10($\downarrow$)\\ \hline
3              &  3.65$\cdot10^{14}      $ &86=27($\uparrow$)+59($\downarrow$) & 19=8 ($\uparrow$)+11($\downarrow$)\\ \hline
1              &  3.65$\cdot10^{14}      $ &80=23($\uparrow$)+57($\downarrow$) & 20=8 ($\uparrow$)+12($\downarrow$)\\ \hline
4 (45$^{0}$)   &  3.65$\cdot10^{14}      $ &77=14($\uparrow$)+63($\downarrow$) & 31=19($\uparrow$)+12($\downarrow$)\\ \hline
\end{tabular}
\caption{Total single bit errors observed  in different registers
of preFPIX2Tb exposed to 200 MeV proton beam
(an upper arrow means a transition from 0 to 1 and a down arrow 
means a transition from 1 to 0).
The observed asymmetry in the DAC upset in the first board
is due to the unequal numbers of zero's (82) 
and one's (30) downloaded into the DAC registers.}
\label{seuerror}
\end{center}
\end{table}
One of the boards (indicated as board 4 in Table ~\ref{seuerror}) 
was placed not orthogonal to the beam, as the other ones, 
but at 45 degrees to explore possible dependence of the error 
rate on the beam incident angle. 
The number of single bit upsets, for an equal amount of total dose, 
is statistically consistent among the various chips. 
In addition, the data do not show any statistically 
significant difference in the error rate between the 
tilted board and the others.
\subsubsection{SEU cross section}
It is common practice to express the error rate in a register as a 
single bit upset cross section $\sigma_{SEU}$, defined as the number 
of errors per bit per unit of integrated fluence: 
\begin{equation}
N_{errors}=F \cdot N_{bit} \cdot \sigma_{SEU}                    
\label{crosssectionformula}
\end{equation}
where N$_{errors}$ is the number of upsets, 
F is the integrated fluence, and
$N_{bit}$ the number of bits exposed.
The single bit upset cross section has been computed for the 
shift-registers and for the DAC registers. 
The results are shown in Table ~\ref{crossection}. 
\begin{table}
\begin{center}
\begin{tabular}{|l|l|l|} \hline
Register                           &  $\sigma_{seu}[10^{-16}cm^{2}]$ \\ \hline
Kill-Inject Mask  (${\uparrow  }$) & 1.0$\pm$0.1		             \\ \hline
Kill-Inject Mask  (${\downarrow}$) & 2.7$\pm$0.2		             \\ \hline
Kill Mask (${\downarrow  }$) clocked at 0.38MHz       & 4.2$\pm$1.2    \\ \hline
DAC                                & 5.5$\pm$0.6		     \\ \hline
\end{tabular}
\caption{Single bit upset cross section for 200 MeV protons 
measured in different registers of preFPIX2Tb 
(an upper arrow means a transition from 0 to 1 and a down arrow 
means a transition from 1 to 0).}
\label{crossection}
\end{center}
\end{table}
Only the statistical error on the cross section has been considered. 
For the shift-registers, the cross section has been computed separately 
for the radiation induced transition from 0 to 1 and from 1 to 0 because 
the data have enough precision to show the existence of an asymmetry. 

The high beam fluence used during the irradiation was of some concern 
regarding any saturation effect in the error rate.  
To study this, we collected some data at a fluence 
of about 4$\cdot$10$^{9}$ protons cm$^{-2}$s$^{-1}$, 
about 5 times less than the nominal fluence. 
In this short test, only one board was irradiated (Apr. '01 test) 
and the single bit cross section was measured to be 
(1.4$\pm$1)$\cdot$10$^{-16}$ cm$^{2}$ 
and (3.5$\pm$1.6 )$\cdot$10$^{-16}$ cm$^{2}$ for the shift-registers 
and (7$\pm$5)$\cdot$10$^{-16}$ cm$^{2}$ 
for the DAC registers in un-clocked mode. This is statistically compatible 
with the results at higher fluence.
\subsubsection{Discussion}
The prediction of the single bit upset cross section 
is very difficult because a lot of parameters 
came into play \cite{Huhtinen}. 
Nevertheless, some gross features of the data can be understood 
simply by some general considerations. 
The disparity in the cross section between the shift registers 
and the DAC registers is likely caused by the different size 
of the active area of the NFET transistors,
which are larger for the DAC register FF's. 
Besides that, the DAC register FF's have a more 
complicated design and an increase in complexity, as a rule of thumb, 
translates to a larger number of sensitive nodes that can be upset.   

The SEU asymmetry for the transition from 0 to 1 with respect to 
1 to 0 can be explained in terms of the FF design. 
The FF's of the shift-registers are D-FF's implemented as 
cross-coupled nor-not gates. Such a configuration has different 
sensitive nodes for 0 to 1 and 1 to 0 upsets. 
No such an asymmetry is expected for the DAC 
registers because the FF's are D-FF's implemented 
as cross-coupled nor-nor gates. 
This symmetric configuration has the distribution of sensitive 
nodes for low logical level the same as when a high logical level is stored.

A decrease of the energy threshold for single bit upset has been reported 
in reference \cite{Faccio} for a static register in clocked mode with 
respect to unclocked mode.  
Our data, taken with a clock frequency of 380 kHz, 
do not show a statistically significant difference from the data 
taken in the unclocked mode.

In reference \cite{O'Neil} a beam angular dependence is expected for 
devices with very thin sensitive volumes that have linear 
energy transfer threshold over 1 MeV cm$^{2}$mg$^{-1}$ and 
tested with 200 MeV protons.  
We didn't observe any dependence of the upset rate on the beam incident angle. 
In fact, due to the smaller device size of the deep submicron elements, 
the sensitive volumes are more cubic than slab shaped.

\subsection{Impact of SEU on BTeV vertex detector}
The vertex detector of BTeV consists of a planar
array of 10 by 10 cm$^{2}$ silicon pixel stations 
arranged perpendicular to the beam and siting inside 
a dipole magnet (about 1.6T). 
Each station is realized with two half L-shaped
substrates staggered along z and covered
by pixel sensors on both sides. The first measuring
accurately one direction perpendicular to the beam (x),
and the second the other direction (y).
The elementary block forming the pixel detector is
an assembly (module) of one long sensor with several readout
chips bump bonded to it and a flex circuit cable which carries
power and signal traces.
The pixel sensor and the FPIX read out chips are the only
active electronics parts on the module.
The FPIX chips are controlled and read out by FPGA's
located outside the magnet.
Thus, the FPIX chips are the only components
which will be susceptible to soft (SEU) errors.
Due to the long interaction region along
the z direction (about 30 cm rms) most of the pixel
detectors are crossed by particles coming
from both directions. 
The insensitivity of the upset rate observed
in our test with respect to the beam incident
angle gives some confidence that the upset cross
section doesn't depend significantly on the direction
of the incoming particle. 
Moreover, the recoiling ions causing upset in the 
electronics have a short range, so 
the material surrounding the readout chips is
unlikely to generate secondary ions
that can increase the upset rate.  
Nevertheless, in future tests we will measure
the SEU cross section for beam incident on the
backside of the chip and with chips bump bonded
to sensors.  

The vertex detector in BTeV is going to be exposed 
mostly to charged hadrons, neutrons, electrons and  
gammas coming from the interaction region.
The particle flux of these species decreases as an inverse
power of the radial distance (the neutron flux reachs a plateau
just outside the vertex detector) and they have a 
broad energy spectrum up to about 100 GeV.   
The upset rate due to the electromagnetic component 
of the radiation field (gammas and electrons) can be neglected,
because they produce few secondary ions in material. 
Table ~\ref{errorrate} shows the expected upset rate due to 
charged hadrons, high energy neutrons, and low energy neutrons
for the different kinds of registers
implemented in the FPIX2 chips.
The cross section values used for the registers 
are the one measured in our tests and should represent
very conservative estimates.
In fact, the SEU upset cross
section for protons saturates at about 200 MeV kinetic energy
and is higher than the cross section for neutrons or pions 
for most of the energy spectrum involved.
The data output serializer has not been implemented yet.
Therefore, in this calculation, a register 24 bits long 
and a SEU cross section equal to 10$^{-15}$cm$^{2}$ 
are assumed for the serializer. We have to keep in mind that
an increase in upset rate in the serializer can be expected
when clocked at high frequency, 
but we don't expect an intolerable rate of soft errors.
The expected SEU bit error rate in the BTeV vertex detector
is small enough that it will not be necessary to design explicitly
SEU tolerant registers.
We believe that the SEU rate can be confortably handled by a periodic
readback of the chip configurations during the data taking 
and a re-download of the chip configuration when an upset
is detected.
\begin{table}
\begin{center}
\begin{tabular}{|l|l|l|l|} \hline
                                   & Ch. Had.	         & n(E$>$14 MeV)    & n(E$<$14 MeV)	\\ \hline
Flux       [s$^{-1}$plane$^{-1}$]  & 1.4$\cdot 10^{8}$   & 1.5$\cdot 10^{7}$& 2.0$\cdot 10^{7}$ \\ \hline
Kill Reg.  [bit $\cdot$ h$^{-1}$]  & 10 		 & 1		    & 1.4		\\ \hline
Inj. Reg.  [bit $\cdot$ h$^{-1}$]  & 27 		 & 2.7  	    & 3.8		\\ \hline
DAC  Reg.  [bit $\cdot$ h$^{-1}$]  & 2  		 & 0.2  	    & 0.3		\\ \hline
Ser. Reg.  [bit $\cdot$ h$^{-1}$]  & 1                   & 0.1  	    & 0.1		\\ \hline
\end{tabular}
\caption{Particle fluxes integrated on one pixel plane in the BTeV experiment
and the predicted single bit error rate for different registers 
(L=$2\cdot10^{32}cm^{-2}s^{-1}$).}
\label{errorrate}
\end{center}
\end{table}
\section{Conclusions}
Total dose and single event tests validate 
the deep submicron CMOS-process FPIX2 designs as radiation tolerant, 
particularly suitable for exposure 
to large integrated total dose.
No evidence of catastrophic failure or deterioration
of the readout chip functionality has been observed
up to dose up of 43 MRad (200 MeV protons).
The single event upset cross sections of static registers 
are relatively small, 
but measurable (from 10$^{-16}$ to 6$\cdot$10$^{-16}$ cm$^{2}$). 
The expected SEU bit error rate in the BTeV vertex detector 
is manageable and does not require further 
SEU hardened registers to be implemented.
Based on the experience gained from the gamma and 
proton irradiation, we intend to submit a 
full-size BTeV pixel readout chip before the end 
of the year 2001. That chip will include the final 
50 micron by 400 micron pixel cells and a high speed 
output data serializer.
\section{Acknowledgements}
We thank Chuck Foster and Ken Murray for the 
generous technical and scientific assistance 
they provided us during the irradiation tests at IUCF.
\noindent
Fermilab is operated by Universities Research Association
under contract with the US Department of Energy.

\end{document}